\documentclass[aps,pra,twocolumn,showpacs,groupedaddress]{revtex4} 

\usepackage{graphicx}  
\usepackage{dcolumn}   
\usepackage{bm}       
\usepackage{amssymb}   
\usepackage{amsmath}

\begin{document}

\title{Pulsed pumping of a Bose-Einstein condensate}

\author{D.\,\,D\"oring}
\author{G.\,R.\,\,Dennis}
\author{N.\,P.\,\,Robins}
\author{M.\,\,Jeppesen}
\author{C.\,\,Figl}
\author{J.\,J.\,\,Hope}
\author{J.\,D.\,\,Close}

\affiliation{Australian Research Council Centre of Excellence for Quantum-Atom Optics, Department of Quantum Science, The Australian National University, Canberra, 0200, Australia}

\date{\today}

\begin{abstract}
In this work, we examine a system for coherent transfer of atoms into a Bose-Einstein condensate.  We utilize two spatially separate Bose-Einstein condensates in different hyperfine ground states held in the same dc magnetic trap. By means of a pulsed transfer of atoms, we are able to show a clear resonance in the timing of the transfer, both in temperature and number, from which we draw conclusions about the underlying physical process.  The results are discussed in the context of the recently demonstrated pumped atom laser.  \end{abstract}

\maketitle 

\section{Introduction}

In 1995, Bose-Einstein condensation in ultracold atomic gases was achieved \cite{Anderson1995,Bradley1995,Davis1995}, an accomplishment that among others led to the invention of the atom laser. Atom lasers are coherent matter waves that bear striking similarities to optical lasers, devices that are nowadays widely used for applications in science, industry and everyday life. The main reasons for the importance of optical lasers are their unique coherence properties and high brightness that offer significant advantages over thermal light sources. In a very similar way, the atom laser is a promising device for use in fields where a high brightness coherent atomic source is required. In the context of high signal-to-noise measurement processes, the achievable brightness of an atom laser may open the route towards unachieved detection senitivities. It is a challenging idea to use an atom laser for atom interferometry and precision measurements. 

Atom lasers are produced by coherently output-coupling an atomic beam from a Bose-Einstein condensate. The atom laser was first realized at MIT \cite{Mewes1997} and has since then been characterized in detail by several groups regarding many of its properties, including the spatial mode profile \cite{Riou2006}, the flux \cite{Robins2006} and the correlation properties \cite{Anton2005}. In order to achieve a truly high brightness and flux in atom lasers, it is crucial to implement a mechanism allowing for continuous operation of the device. So far, the average flux of an atom laser has been limited by the repetition cycle of the apparatus producing the Bose-Einstein condensate. An atom laser can only be output-coupled until the Bose-Einstein condensate that serves as a source (the lasing condensate) is depleted. For continuous operation it is necessary to implement a mechanism that coherently replenishes the lasing condensate.

\section{Pumping a Bose-Einstein condensate} 

Recently, our group has achieved such a pumping mechanism in the regime where the replenishment is realized at time scales corresponding to quasi-continuous operation of the atom laser (of the order of $100\,\text{ms}$) \cite{Nicholas2008}. In this system a continuous atom laser beam is produced in the $\left|F=2,m_F=0\right\rangle$ hyperfine ground state of $^{87}\text{Rb}$ from a trapped  $\left|F=2,m_F=2\right\rangle$ condensate using radio frequency (rf) induced spin-changing transitions. The falling atoms are exposed to a continuous, upward propagating light field with a well-defined linear polarization.  After a short fall distance, the atoms in the $\left|F=2,m_F=0\right\rangle$ state enter a large $\left|F=1,m_F=-1\right\rangle$ condensate.  The measurements show that up to 35\% of the original $\left|F=2,m_F=2\right\rangle$ cloud is coherently transferred to the $\left|F=1,m_F=-1\right\rangle$ condensate, for an operation time of $200\, \text{ms}$. The
\begin{figure}[b]
\includegraphics[scale=0.9]{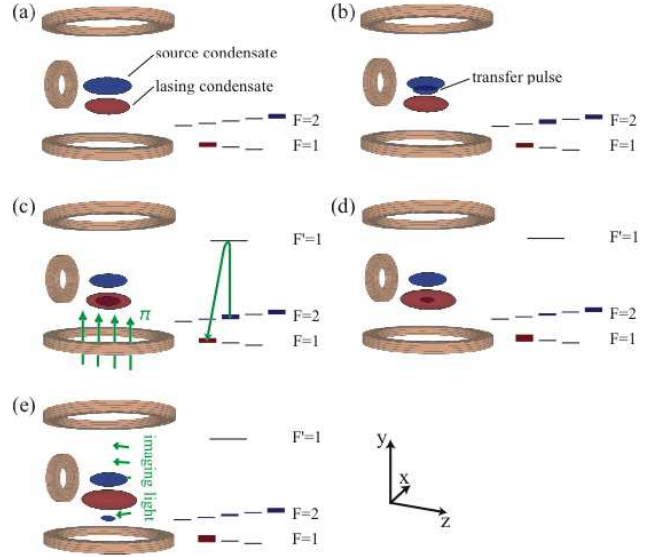}
\caption{\label{fig:scheme}(Color online) Sequential illustration of the experiment. The setup includes a source and a lasing condensate trapped in the same QUIC-trap (a). A pulse of atoms is output-coupled from the source cloud and is accelerated downward due to gravity (b). As the atoms fall, a variable delay pulse of $\pi$-polarized light is applied (c). The atoms in the transfer pulse continue to fall and are detected via absorption imaging (e).}
\end{figure}
inferred mechanism for this transfer is that the $\left|F=1,m_F=-1\right\rangle$ condensate stimulates transitions from the $\left|F=2,m_F=0\right\rangle$ state mediated by the linearly polarized optical field, effectively `drawing' them in from the falling beam. This implies that the momentum of the transferred atoms has to vanish in order to match the (zero) momentum of the pumped condensate. The possible resonances corresponding to this matching condition relate to a momentum transfer of either $0\hbar k$ or $2\hbar k$ from the light field to the atom laser beam, where $\hbar k$ is the momentum of the photons driving the pumping process. As discussed in \cite{Nicholas2008}, the two possible underlying processes for the stimulated transition are (a) Bose-stimulated Raman scattering \cite{Yutaka2004,Dominik2004} and (b) resonant coupling driven by electromagnetically-induced transparency (EIT) \cite{Naomi2007}. Each of these two processes has characteristic features regarding the above mentioned momentum resonance. Whereas in case (a) both of the momentum resonances are in principle conceivable, the process analogous to the work in \cite{Naomi2007} is restricted to a momentum transfer of $0\hbar k$. By investigating the occurrence of momentum resonances, one can draw conclusions about the underlying physical process of the pumping. The results from our previous work clearly showed coherent pumping, but the continuous nature of the experiment and the extreme stability required to operate it made it difficult to make quantitative measurements of the temporal dependence and the above mentioned momentum resonances. 

\begin{figure}[b]
\includegraphics[scale=0.73]{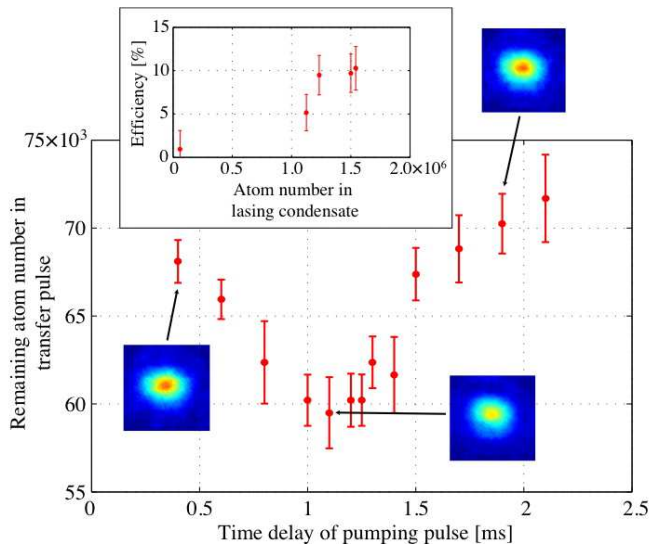}
\caption{\label{fig:graph}(Color online) Remaining atom number in the transfer pulse as a function of the time delay of the pumping pulse. The absorption pictures show the remaining atoms in the transfer pulse after the pumping process. The inset depicts the efficiency of the pumping process as a function of the size of the lasing condensate.}
\end{figure}
\begin{figure}[bht]
\includegraphics[scale=0.87]{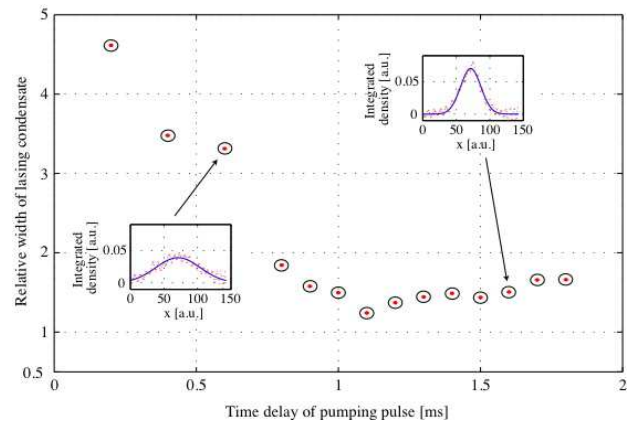}
\caption{\label{fig:heating}(Color online) Width of the lasing condensate (normalized to the width with no pumping light applied) as a function of the time delay of the pumping pulse. The two insets show the integrated density profile of the (expanded) lasing condensate for two different time delays of the pumping pulse.}
\end{figure}

Our aim in this work is to isolate and measure the process that drives the pumped atom laser. To do so we take a step backward in complexity, and  present experimental and theoretical results on this pumping mechanism operating in the pulsed regime. A coherent population transfer between a source and a lasing condensate is realized by means of an atom laser transfer pulse. The timescales of the population transfer are of the order of the frequency width of the condensates. This offers the opportunity to characterize the pumping mechanism in a different temporal regime and to use a different detection channel on the underlying process. As opposed to the work in \cite{Nicholas2008}, we detect the population transfer by measuring the depletion of the transfer pulse instead of an increase of atom number in the lasing mode after the pumping. Additionally, we measure the temperature of the lasing condensate after the pumping pulse and present data on the energy spectrum of the two-condensate system.   

\section{Experimental setup}

A detailed description of the experimental setup is given in \cite{Daniel2008}. We produce two Bose-Einstein condensates in the hyperfine ground states $\left|F=1,m_F=-1\right\rangle$ and $\left|F=2,m_F=2\right\rangle$. The two atomic clouds are trapped in the same harmonic magnetic trap and are spatially separated by $7.3\,\mu \text{m}$ due to their different gravitational sag. The upper (lower) of the two condensates is henceforth referred to as the source (lasing) condensate. 
By radio frequency (rf) induced spin-flips we can coherently transfer atoms into a magnetically insensitive state and thus output-couple an atom laser beam from either of the two clouds.  For the implementation of the pulsed pumping mechanism (see Fig. \ref{fig:scheme}), we apply a short ($40\,\mu \text{s}$) output-coupling pulse to the source condensate. The $\left|F=2,m_F=0\right\rangle$ output-coupled pulse of atoms (the transfer pulse) accelerates downward due to gravity, and propagates through the $\left|F=1,m_F=-1\right\rangle$ condensate. At a well defined time during the travel of the transfer pulse, we apply a pulse ($150\,\mu\text{s}$) of linearly polarized light in the vertical direction, opposite to the movement of the atoms as they fall under gravity. The intensity of this pumping light is of the order of $30\, \mu\text{W/cm}^2$ and is adjusted such that for the pulse length chosen we observe a loss of $\sim 20\,\%$ from the transfer pulse due to spontaneous emission. Thus we make sure that there is a reasonable amount of interaction between the transfer pulse and the pumping light which can then be significantly increased by Bose-stimulation in presence of the lasing mode. The light is blue-detuned by two natural linewidths from the $F=2 \rightarrow F'=1$ transition. In order to investigate the pumping process, we adjust the timing of the pumping light pulse applied to the system.   

\section{Results and discussion}

Figure \ref{fig:graph} displays the number of atoms left in the transfer pulse after the pumping when we vary the time delay of the light pulse. Changing the pulse delay clearly affects the atom number remaining in the transfer pulse. The data shows a resonance centered at a time $1.1\,\text{ms}$ after the output-coupling pulse, suggesting an enhanced transfer of atoms into the lasing condensate. In Fig. \ref{fig:heating} we show the heating of the $\left|F=1,m_F=-1\right\rangle$ cloud as a function of the delay time, measured with a $100\,\mu\text{s}$ long pumping light pulse. Again the curve reaches a minimum value with no significant heating at $1.1\,\text{ms}$. The heating does not increase for delay times above the center of the resonance, which may be explained by the transfer pulse shielding the lasing condensate from the emitted resonant photons. The sets of data in Fig. \ref{fig:graph} and Fig. \ref{fig:heating} indicate that the underlying process occurs predominantly at a pulse delay time of $1.1\,\text{ms}$, and the resonance data shows that the $\left|F=1,m_F=-1\right\rangle$ condensate mediates the process, since without this condensate there is no loss from the pulse (see inset in Fig. \ref{fig:graph}).  

The $1.1\,\text{ms}$ resonance is consistent with the time a free falling atom takes to travel the distance of $7.3\,\mu\text{m}$ between the two condensate centers ($1.2\,\text{ms}$). In this time the atoms reach a velocity corresponding to $1.3\,\text{cm}/\text{s}$. This velocity is equivalent to the absorption and subsequent emission of a photon, giving a $2\hbar k$ momentum kick. Indeed, we set the distance between the $F=1$ and $F=2$ condensates in order to achieve exactly this situation.  Our measurements support this, indicating that the optical pumping pulse effectively stops the atoms. This requires that the emitted photon must be emitted almost straight down and shows that the emission process is not spontaneous.

\begin{figure}[b]
\includegraphics[scale=0.58]{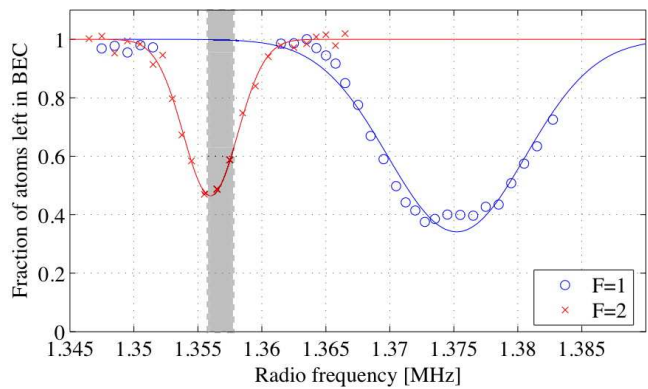}
\caption{\label{fig:spectroscopy} (Color online) Spectroscopic results for the source ($F=2$, red crosses) and the lasing ($F=1$, blue circles) condensate. The solid lines are Gaussian fits to the data. The operating region (shaded area) is clearly separated from the overlap region between the two condensates.}
\end{figure}

To prevent a disturbance of the pumping process by the atom laser output-coupling, the measurements presented in Fig.~\ref{fig:graph} and Fig.~\ref{fig:heating} do not extend to a zero time delay. The data in Fig.~\ref{fig:heating}, however, suggests a large heating rate when approaching small time delays. More information on the zero time delay resonance corresponding to a vanishing momentum transfer is obtained from a spectroscopic analysis of the source and the lasing condensate (Fig.~\ref{fig:spectroscopy}). We apply a $100\,\text{ms}$ weak radio frequency pulse to the system and separately measure the loss from each of the two condensates due to this output-coupling process. The measurement is repeated for different output-coupling radio frequencies, effectively addressing different spatial regions of the two-condensate system. Using this state-selective data on the loss from the two condensates, we create resonance loss curves as displayed in Fig.~\ref{fig:spectroscopy}. The operating region for the long pulse atom laser in the pumping experiment \cite{Nicholas2008} is indicated by the shaded area. The area is clearly separated from the small overlap region of the source and the lasing condensate. This overlap region is the only place where an EIT-like pumping process with vanishing momentum transfer could occur. Due to the narrow spectral width of the output-coupling field, this process can therefore be ruled out for the continuous pumping in \cite{Nicholas2008}.  

From the data presented above, we draw three main conclusions: (a) The $\left|F=1,m_F=-1\right\rangle$ condensate significantly enhances transitions of atoms from the transfer pulse into the $\left|1,-1\right\rangle$ state. (b) These transitions primarily occur at a position in space that maximally overlaps with the lasing condensate. (c) This position allows the absorption of a pump beam photon and emission of a photon downward that exactly cancel the $2\hbar k$ momentum gained in falling to that position. The latter point includes that the described $2\hbar k$ momentum resonance is favorable compared to a $0\hbar k$ resonance, where atoms are transferred immediately after the output-coupling of the transfer pulse in the (very small) overlap region of the two condensates. As mentioned above, this implies that the underlying mechanism for the pumping is a Raman superradiance-like process and not a resonant coupling driven by EIT. 

\section{Theoretical model}

\begin{figure}
\includegraphics[scale=0.5]{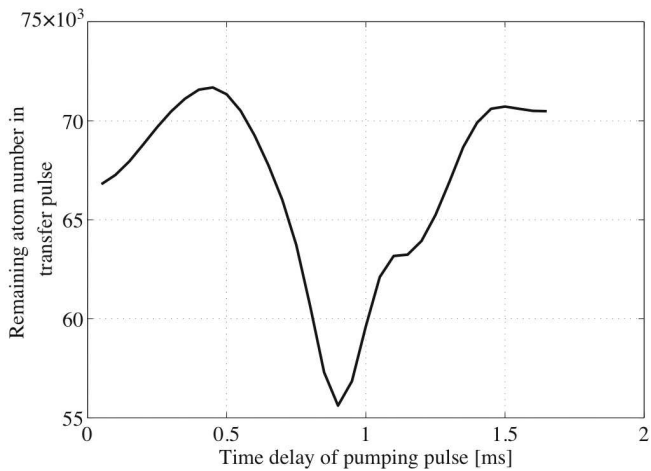}
\caption{\label{fig:theory} Theoretical results for the remaining atom number in the transfer 
pulse as a function of the time delay of the pumping pulse.}
\end{figure}

To confirm these conclusions, we have modelled our experiment theoretically using the Maxwell-Schr\"{o}dinger equations \cite{Zobay2006} which are a mean-field approximation to the coupled light-atom system derived from the Hamiltonian
\begin{equation}
\hat{H} = \hat{H}_\text{atom} + \hat{H}_\text{rf} + \hat{H}_
\text{light} + \hat{H}_\text{int},
\end{equation}
with
\begin{equation}
\hat{H}_\text{int} = - \sum_{g, e} \int \left(\hat{\Psi}^
\dagger_e(\mathbf{r}) \hat{\Psi}_g(\mathbf{r}) \mathbf{d}_{eg} + h.c. 
\right) \cdot \hat{\mathbf{E}}(\mathbf{r})\, d^3\mathbf{r}.
\end{equation}
$\hat{H}_\text{atom}$ includes the usual kinetic energy, trapping potential and mean-field interaction terms for the $F=1, 2$ manifolds, $\hat{H}_\text{rf}$ is the rf output-coupling term and $\hat{H}_\text{light}$ is the self-energy of the light field. The electric dipole moment between states $e$ and $g$ is $\mathbf{d}_{eg}$, and $\hat{\mathbf{E}}(\mathbf{r})$ is the electric field operator. In this model, the mean-field interaction has been retained, and spontaneous emission has been approximated as a loss term for the appropriate atomic states. As shown in Fig.~\ref{fig:theory}, the calculated change in atom number of the transfer pulse is in quantitative agreement with the experimental results in Fig.~\ref{fig:graph}.  In addition to the expected minimum in Fig.~\ref{fig:theory} near a $1\, \text{ms}$ delay, there is a local minimum at zero delay corresponding to the $0\hbar k$ momentum resonance. A comparison to the heating curve in Fig.~\ref{fig:heating} and the spectroscopy in Fig.~\ref{fig:spectroscopy} however excludes the possibility that the $0\hbar k$ momentum resonance could have significantly contributed to the pumping effect observed in our previous continuous experiment \cite{Nicholas2008} and further investigated in this paper. The occurrence of the $0\hbar k$ momentum resonance in Fig.~\ref{fig:theory} can be attributed to the Fourier width of the output-coupling pulse, that is comparable to the spectral width of the source condensate and allows output-coupling in the overlap region of the two clouds. Due to its mean-field nature, our theory does not allow for a calculation of heating so that we cannot model the results in Fig.~\ref{fig:heating}.

The heating measurements of the lasing condensate as a function of the pumping delay display a feature that one would not intuitively expect. At a delay of approximately $1.1\, \text{ms}$ there is an increase in the number of atoms lost from the transfer pulse as shown in Fig.~\ref{fig:graph}, and this must correspond to an increase in the number of photons emitted that are resonant with the lasing condensate. Despite this, there is significantly less heating of the lasing condensate when the pumping pulse is applied at a $1.1\, \text{ms}$ delay than there is when it is applied at, for example, a delay of $0.4\, \text{ms}$ (see Fig.~\ref{fig:heating}), when fewer resonant photons are emitted.  This reduction in heating cannot be explained by collisional heating from the falling atoms. Although each falling atom will at most have $1.4$ times the chemical potential of the lasing condensate in kinetic energy, the atom number in the transfer pulse is only $20\,\%$ of the number in the lasing condensate. One possible explanation for this observed reduction in spontaneous heating is a quantum-mechanical destructive interference between two heating processes. This possibility has been shown theoretically to lead to a suppression of spontaneous heating in the boson-accumulation regime (BAR) \cite{Cirac1996,Floegel2001}. The experiment described here is deep into the BAR which requires that an excited atom is more likely to be stimulated to decay into the condensate mode than to emit a photon in a random direction. For our experimental parameters, we estimate this branching ratio to be $\eta \approx 100$.

\section{Conclusion}

The work presented in this paper demonstrates coherent population transfer between two Bose-Einstein condensates. By operating in the pulsed regime, this experiment permits the temporal investigation of the momentum resonance required for the population transfer. Due to the slight overlap of the two condensates, two momentum resonances were possible. The atoms in the transfer pulse (or beam) could absorb the optical pumping light shortly after being output-coupled and emit a photon in the same direction to decay into the mode of the lasing condensate with no change in momentum. Alternatively, the atoms could absorb the optical pumping light approximately $1.2\,\text{ms}$ after output-coupling when they have a momentum of $2\hbar k$ downward, and emit a photon in the opposite direction to decay into the mode of the lasing condensate. It was not possible to distinguish between these two in our previous experiment demonstrating a continuously pumped atom laser \cite{Nicholas2008}. From the results presented in this work, we conclude that the emitted photons are emitted downward corresponding to a $2\hbar k$ momentum kick given to the atoms. Using the information about this momentum resonance, we conclude that it is a Raman superradiance-like process that drives the pumping of the atom laser. One could expect the emitted resonant photons to significantly heat the condensate due to re-absorption and spontaneous emission. However, we find no observable increase in the temperature of the condensate. We have reason to believe that this is due to effects beyond mean-field theory such as those that apply in the boson-accumulation regime \cite{Cirac1996}. These effects are not negligible and must be considered in a full theoretical treatment. Further investigations of this effect are planned to be the subject of future experiments. 

\section*{Acknowledgments}

This work was financially supported by the Australian Research Council Centre of Excellence program. Numerical simulations were done at the NCI National Facility.

\end{document}